\documentclass[fleqn,10pt]{wlscirep}
\usepackage[utf8]{inputenc}
\usepackage[T1]{fontenc}
\usepackage{comment}
\usepackage{booktabs}

\usepackage{subcaption}
\usepackage{algpseudocode}
\usepackage{algorithm}

\usepackage{hyperref, bookmark}
\hypersetup{
  pdftitle={Routing and Scheduling Optimization for Urban Air Mobility Fleet Management using Quantum Annealing},
  pdfauthor={Renichiro Haba, Takuya Mano, Ryosuke Ueda, Genichiro Ebe, Kohei Takeda, Masayoshi Terabe, Masayuki Ohzeki},
  pdfkeywords = {optimization, quantum annealing, QUBO, urban air mobility, centralized, routing algorithm}
}

\usepackage{color}

\title{Routing and Scheduling Optimization for Urban Air Mobility Fleet Management using Quantum Annealing}

\author[1,2,*]{Renichiro Haba}
\author[3]{Takuya Mano}
\author[3]{Ryosuke Ueda}
\author[3]{Genichiro Ebe}
\author[3]{Kohei Takeda}
\author[3]{Masayoshi Terabe}
\author[1,2,4]{Masayuki Ohzeki}

\affil[1]{Graduate School of Information Sciences, Tohoku University, Sendai, Japan}
\affil[2]{Sigma-i Co., Ltd., Tokyo, Japan}
\affil[3]{Sumitomo Corporation, Tokyo, Japan}
\affil[4]{Department of Physics, Tokyo Institute of Technology, Tokyo, Japan}

\affil[*]{renichiro.haba.r6@dc.tohoku.ac.jp}

\begin{abstract}
\addcontentsline{toc}{section}{Abstract}
The growing integration of urban air mobility (UAM) for urban transportation and delivery has accelerated due to increasing traffic congestion and its environmental and economic repercussions. 
Efficiently managing the anticipated high-density air traffic in cities is critical to ensure safe and effective operations.
In this study, we propose a routing and scheduling framework to address the needs of a large fleet of UAM vehicles operating in urban areas.
Using mathematical optimization techniques, we plan efficient and deconflicted routes for a fleet of vehicles. 
Formulating route planning as a maximum weighted independent set problem enables us to utilize various algorithms and specialized optimization hardware, such as quantum annealers, which has seen substantial progress in recent years.
Our method is validated using a traffic management simulator tailored for the airspace in Singapore. 
Our approach enhances airspace utilization by distributing traffic throughout a region.
This study broadens the potential applications of optimization techniques in UAM traffic management.
\end{abstract}
\begin{document} 

\flushbottom
\maketitle

\thispagestyle{empty}

\section*{Introduction}
\addcontentsline{toc}{section}{Introduction}
Increasing city traffic induces heavy congestion, resulting in immense economic costs and adverse environmental impacts every year. 
To address the issue, urban air mobility (UAM) systems have been envisioned as a technology that utilizes automated air transportation services to carry people or cargo at lower altitudes in and around metropolitan areas \cite{straubingerOverviewCurrentResearch2020}. 
As the demand for UAM operations increases, there is a growing need for a systematic approach to manage their traffic, especially in low-altitude airspace. 
Recent NASA-commissioned market studies estimate that by 2030, there could be up to 500 million flights annually for package delivery services and 750 million flights for air metro services, making UAM a highly profitable and viable enterprise\cite{gipsonUAMOverview2019}.
Consequently, this poses a major challenge for aviation authorities and air navigation service providers as they attempt to integrate these new and novel operations into the national airspace.
Unmanned aircraft systems traffic management (UTM) are conceptual frameworks for the safe and efficient management of the enormous number of aircraft anticipated above cities and people\cite{UTMConceptOperations2020}.
A primary goal is to ensure the safe separation of UAM vehicles from each other and from other airspace users, such as traditional manned air traffic. 
This can be done through strategic deconfliction and dynamic scheduling of UAM flight requests during flight planning. 
However, while in flight, unforeseen contingencies, such as weather events, emergencies, or infrastructure outages, may require a UAM vehicle to dynamically change its route to avoid the contingency. 
In high-density operations, this change in route will cause cascading conflicts for other active operations. 
To help maintain safe airspace operation, preflight strategic deconfliction and in-flight tactical deconfliction are critical \cite{prevotUASTrafficManagement2016a}.

In route planning with strategic deconfliction, flight paths are designed before launch based on demand, considering factors such as traffic density, aerodrome capacity, weather conditions, and both permanent and temporary flight restrictions.
Routing is widely applied in various fields, including shortest route suggestions in-car navigation systems and the control of automated guided vehicles (AGVs) in factories and warehouses\cite{duckhamSimplestPathsAutomated2003, ohzekiControlAutomatedGuided2019, habaTravelTimeOptimization2022a}.
Standard routing algorithms include sampling-based methods like rapidly-exploring random trees (RRT) \cite{lavalleRapidlyexploringRandomTrees1998} and exact algorithms such as Dijkstra’s algorithm \cite{dijkstraNoteTwoProblems1959} and A-star (A*) \cite{hartFormalBasisHeuristic1968}.
However, these algorithms are not directly applicable to UAM routing, as they do not account for collision avoidance.
Routing algorithms with collision avoidance have been actively studied in the field of unmanned aerial vehicles (UAVs) rather than UAM vehicles\cite{coutinhoUnmannedAerialVehicle2018, fugenschuhFlightPlanningUnmanned2021a, forsmoOptimalPathPlanning2012}.
Most studies on UAV route design assume that UAVs can move freely in three-dimensional space. In contrast, only a few studies consider route design within urban environments, where the available airspace for flight is significantly restricted.

According to the concept of operations (ConOps) for UTM, urban air space is modeled by a comprehensive routing network composed of nodes and lines in the real space to mitigate the complexity of operations \cite{mohamedsallehConceptOperationsConOps2017a}.
Under such a condition, strategic deconfliction methods have been proposed, and take-off and landing time for each vehicle is controlled \cite{daiConflictfreeFourdimensionalPath2021b, sacharnyLaneBasedLargeScaleUAS2022}.
Efficient fleet management remains a significant challenge, particularly in cities where numerous flight operations occur within constrained airspace.
The complexity of managing such operations increases as the number of flights grows, requiring sophisticated coordination to avoid congestion and delays.
One approach to this challenge is mathematical optimization, which can help identify the best solutions under defined objectives and constraints.
Fleet control under some objective to be minimized or maximized can be translated into mathematical optimization problems, which are usually hard to solve optimally for its NP-hardness as represented by vehicle routing problem \cite{dantzigTruckDispatchingProblem1959}. 

Metaheuristics, which are general-purpose algorithms for a wide range of problems, have attracted a great deal of attention in recent years, such as genetic algorithms and simulated annealing \cite{hollandAdaptationNaturalArtificial, kirkpatrickOptimizationSimulatedAnnealing1983}. 
Furthermore, in recent years, methods utilizing quantum nature have emerged, and both industry and academia are actively attempting to transcend classical computation.
Quantum annealing is a metaheuristic for combinatorial optimization problems and utilizes quantum fluctuations for global search\cite{kadowakiQuantumAnnealingTransverse1998a}.
The method is specialized in solving quadratic unconstrained binary optimization (QUBO) problems, and many well-known problems can be translated into \cite{lucasIsingFormulationsMany2014b}.
In the ideal procedure, quantum annealing outputs the optimal solution by slowly decreasing the strength of the fluctuation of binary variables. 
The quantum adiabatic theorem ensures that the ground state, which corresponds to the optimal solution, is obtained by evolving the system adiabatically \cite{suzukiResidualEnergiesSlow2005a, morita_mathematical_2008, ohzekiQuantumAnnealingIntroduction2011a}.
The hardware implementing quantum annealing developed by D-Wave Systems, Inc. have become commercially available, marking a significant milestone in practical applications of quantum annealing.
In contrast to theoretical aspects, the machines do not perform quantum annealing ideally, and their optimization performance is quite limited at the present stage.
However, the rapidness of sampling can be effective for attaining relatively good solutions as a heuristic solver.
To inspect the industry applicability of D-Wave's quantum annealer, a wide variety of practical use cases have been explored in finance\cite{rosenbergSolvingOptimalTrading2016a, orusForecastingFinancialCrashes2019, venturelliReverseQuantumAnnealing2019}, traffic\cite{neukartTrafficFlowOptimization2017b, hussainOptimalControlTraffic2020a, shikanaiTrafficSignalOptimization2023}, logistics\cite{feldHybridSolutionMethod2019a}, manufacturing\cite{ohzekiControlAutomatedGuided2019, habaTravelTimeOptimization2022a, venturelliQuantumAnnealingImplementation2016a}
, and marketing\cite{nishimuraItemListingOptimization2019}, as well as in decoding problems\cite{ideMaximumLikelihoodChannel2020, araiMeanFieldAnalysis2021b}.
In this study, we use the metaheuristic optimization method for fleet management.
Specifically, we realize the route planning by dynamically solving maximum weighted independent set (MWIS) problems, which are well-known graph theoretical problems. 

The remainder of this paper is organized as follows: In the next section, we outline the fundamental principles of fleet management. 
We introduce our routing and scheduling framework, which leverages optimization techniques, and explain how the problem is structured. 
In the following section, we evaluate our method through the UTM simulator developed by OneSky Systems.
To validate the effectiveness of quantum annealing, we compare the results of a greedy algorithm and a classical commercial optimizer.
In the final section, we summarize our findings and discuss potential avenues for future research to enhance UAM traffic management.

\section*{Method}
\addcontentsline{toc}{section}{Method}
In this section, we introduce our routing and scheduling method for the UAM vehicles fleet.

\begin{figure}[tbp]
  \centering
  \includegraphics[width=14cm]{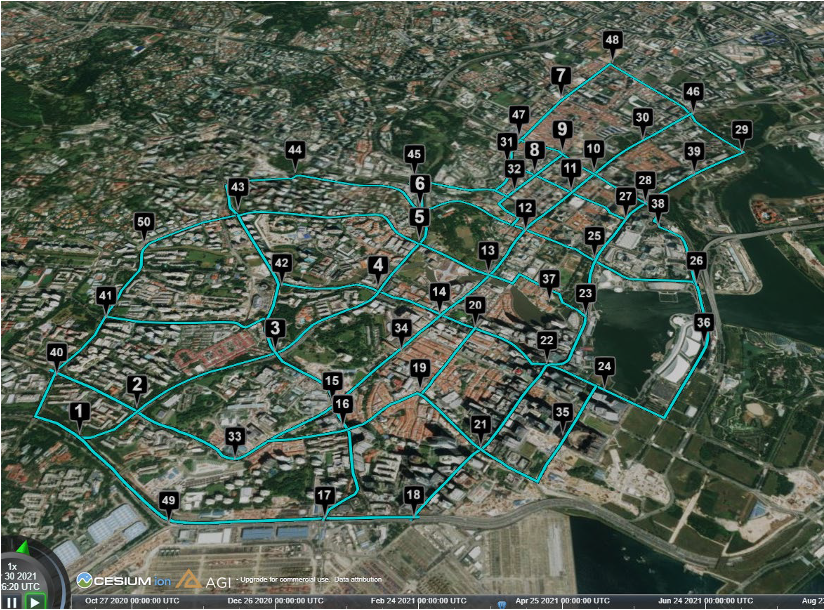}
  \caption{\textbf{Routing Network in Singapore.} The labeled nodes represent aerodromes, while the lines represent corridors. This airspace structure was generated using the OneSky UTM simulator.}
  \label{fig:routing_network}
\end{figure}

To manage urban airspace, aerodromes, and flight corridors are designed based on the surrounding environment, including artificial structures and geographical features. 
The UAM vehicles can take off, land at these aerodromes, and navigate designated corridors.
As the airspace is to be integrated with the existing urban infrastructures, to ensure the safety of other users, UAM vehicles will fly past as little human activity area within the urban space as possible, such as the roof of buildings, canals, and drainage systems \cite{prevotUASTrafficManagement2016a}.
Consequently, in urban environments, the available flyable space is significantly constrained, and the airways for the UAM vehicles can be modeled as a graph-structured routing network.
An example of a routing network within Singapore's urban environment is illustrated in Fig. \ref{fig:routing_network}.
Flight requests are submitted on demand to transport cargo or passengers. 
Each request consists of a pair of nodes representing the take-off and landing locations within the routing network. 
Additionally, requests can include details such as the desired start time or the latest allowable start time.

Given a set of flight requests, the goal of routing and scheduling is to determine, as an output, the start time and flight path for each request. 
The start time refers to the take-off time and must adhere to the time-window constraints defined by the desired and allowable start times. 
The flight path consists of a sequence of nodes and edges, representing the movement of the UAV from the source to the destination within the routing network.
As urban flyable airspace is quite limited, routing and scheduling must be managed with enough efficiency to handle the high volume of anticipated flight requests.
We break down the efficiency of management with the number of approved requests per time unit and the shortness of determined flight path lengths.
Approving as many requests as possible creates flexibility for future requests by opening up the room for potentially incoming demands, while keeping flight paths short helps avoid wasting both flight time and airspace.
To maximize overall efficiency, the start times and flight paths should be adjusted simultaneously for multiple requests, rather than processing them individually.

For the safety of flight operations, UAM vehicles must maintain sufficient separation from one another by adhering to minimum distance requirements in the airspace.
The concept of strategic deconfliction ensures that flight paths are calculated to avoid collisions with other aircraft, both in-flight and scheduled. 
To achieve this, the precise position of each UAM vehicle is simulated over time by following its kinematic motion, and the time-dependent distance between any two vehicles is continuously monitored.
Flights are approved only if this distance remains greater than the minimum required separation.
In practice, re-routing, in addition to pre-flight deconfliction, must also account for dynamic changes in airspace, such as geofence restrictions.
For simplicity, we assume that the airspace remains fixed during operation, with no dynamic changes in the flyable area.

\begin{figure}[tbp]
  \centering
  \includegraphics[width=15cm]{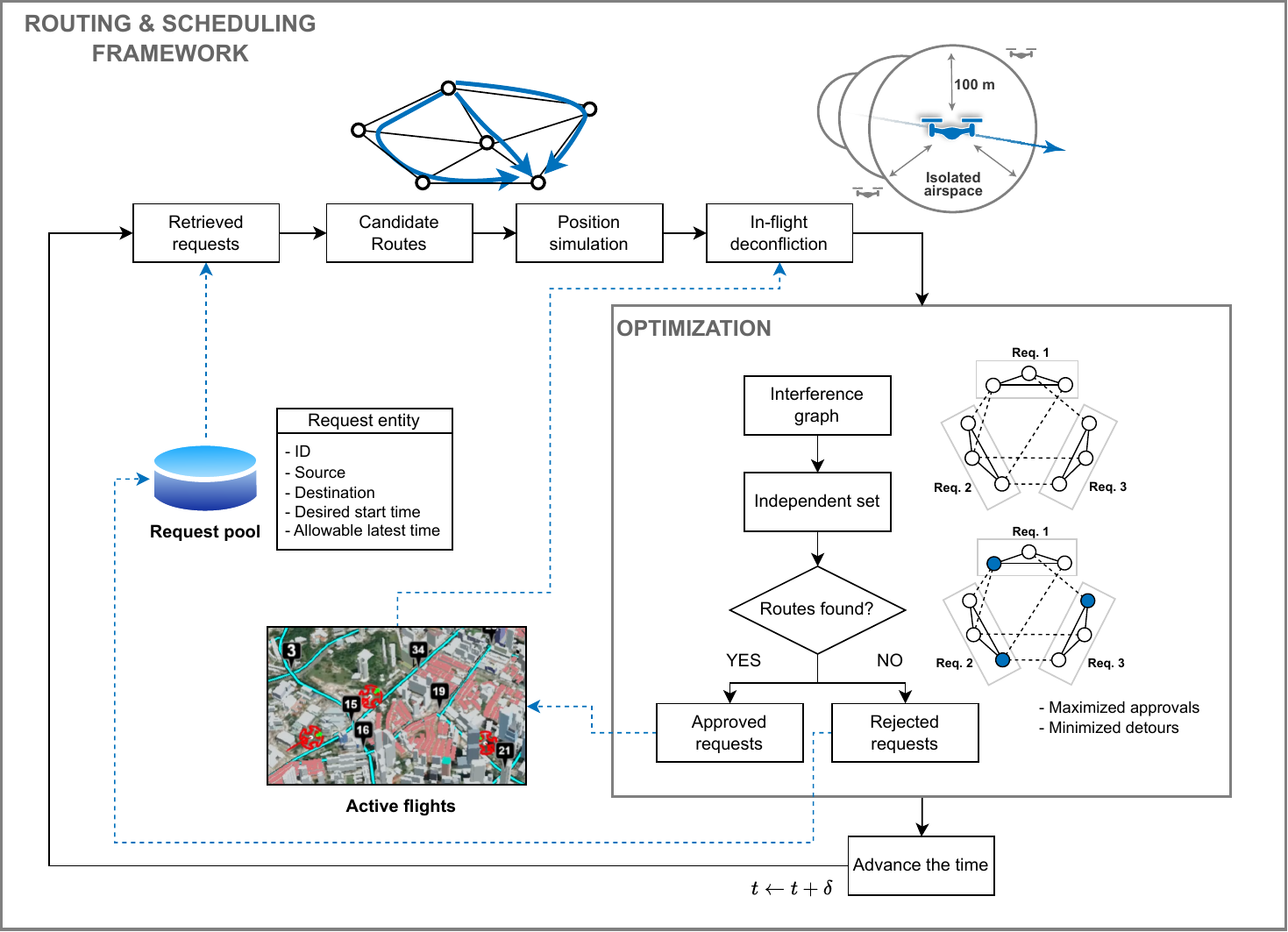}
  \caption{\textbf{Dynamic scheduling and routing framework.} The framework consists of two main components: route generation, and optimization.}
  \label{fig:framework}
\end{figure}

In this study, we introduce a dynamic scheduling and routing framework to efficiently process a high volume of dense flight requests.
The framework is depicted in Fig. \ref{fig:framework}.
The framework operates through the following procedure:
\begin{enumerate}
  \item Gather information on requests that are eligible to begin flights at time $t$.
  \item For each request, generate a set of possible candidate routes.
  \item Simulate the positions of all candidate routes over time.
  \item Calculate the distance between each pair of candidate routes and any active flight routes. Discard candidate routes if the separation distance falls below the minimum required threshold.
  \item For distinct requests, calculate the distance between each pair of candidate routes at each time step. 
  \item Select the best routes that satisfy the deconfliction constraints using an optimization method.
  \item Schedule each request with a valid route as a flight starting at time $t$.
  \item Increment time by the defined time interval and repeat the process from step 1 until the operation concludes.
\end{enumerate}
The details of specific components are described as follows.
In the process of generating candidate routes, a set of flight paths is generated for each request, from which one will be selected as the scheduled flight route.
It is desirable that the candidate routes are not only short but also sufficiently distinct from one another to increase the flexibility for system-wide adjustments.
To achieve this, a Dijkstra-based algorithm is employed to generate a diverse set of short routes.
In the standard Dijkstra algorithm, the distances of edges are used as weights to calculate the shortest path.
In our method, once the shortest path is determined using Dijkstra’s algorithm, a penalty is applied to the edge weights of the routing network that the path traverses, and the process is repeated to compute additional routes.
This penalization helps avoid outputting similar routes and, as subsequent paths are less likely to overlap with the previous ones.
Through empirical testing, we have found that applying a penalty equal to five times the original edge weight effectively generates a sufficient variety of dissimilar routes.
After the candidate routes are generated, each route is translated in to detailed flight paths by simulating the position of UAM vehicles at each time step.
For simplicity, we assume that the UAM vehicles fly at a constant speed, and their positions at each time step are calculated using linear interpolation.
At each time step throughout the journey, the distance between the UAM vehicles' position and other active flight routes is computed. If the separation distance falls below the minimum required threshold, the candidate route is discarded.
To assess potential interference between candidate routes, the distance between each pair of candidate routes for distinct requests is calculated at each time step. 
If the separation distance falls below the required minimum threshold, the pair of candidate routes is considered to be in conflict. 
In such cases, no more than one route from the conflicting pair can be scheduled as a flight.
Based on this interference information, an optimization method is applied to select the most suitable routes that satisfy the separation constraints.

We employ mathematical optimization techniques to determine the optimal combination of routes.
In optimization, a problem is formulated as the minimization or maximization of an objective function subject to constraints.
Specifically, we formulate the route selection process as a maximum weighted independent set (MWIS) problem, which is a well-studied combinatorial optimization problem.
In the MWIS problem, a graph $G=(V,E)$ is given, along with a weight $w_i$ for each vertex $i \in V$.
The objective is to find a subset $I \subseteq V$ such that no two vertices in $I$ are adjacent, and the sum of the weights of the vertices in $I$ is maximized\cite{sanghaviMessagePassingMaximum2009,kakoApproximationAlgorithmsWeighted2005, trevisanInapproximabilityCombinatorialOptimization2014, sakaiNoteGreedyAlgorithms2003b}.
We denote the set of vertices and edges in $G$ as $V(G)$ and $E(G)$, respectively.
The neighborhood of a vertex $i$ in graph $G$ is denoted as $N(i)$, and the degree of vertex $i$ is denoted as $d(i)$.

The MWIS problem can also be formulated as integer programming problem as follows:
\begin{equation}
\begin{split}
  \text{maximize} & ~~~~ \sum_{i \in V}  w_i x_i \\
  \text{subject~to} & ~~~~ x_i + x_j \leq 1 ~~~~ \forall ij \in E \\
  & ~~~~ x_i \in \{0,1\} ~~~~ \forall i \in V.
  \label{eq:MWIS_IP}
\end{split}
\end{equation}

\begin{figure}[tbp]
  \centering
  \begin{subfigure}[t]{0.45\linewidth}
    \includegraphics[width=\linewidth]{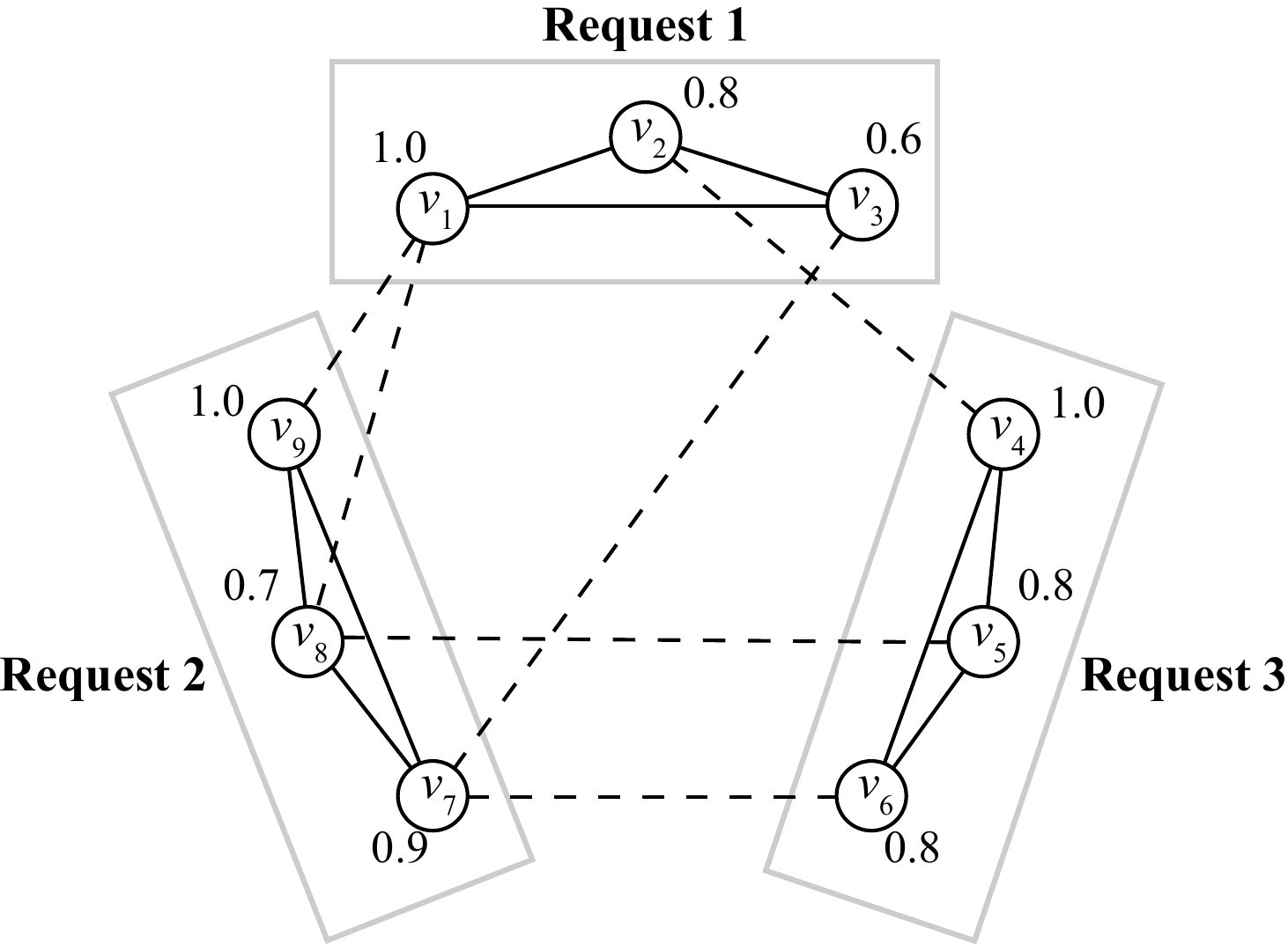}
    \caption{\textbf{A generated graph.} Each vertex represents a candidate route. Solid and dashed edges represent interference between candidate routes for the same request and conflicts between candidate routes for different requests, respectively. The numbers on the vertices represent their corresponding weights.}
    \label{fig:MWIS_graph_instance}
  \end{subfigure}
  \hfill
  \begin{subfigure}[t]{0.45\linewidth}
    \includegraphics[width=\linewidth]{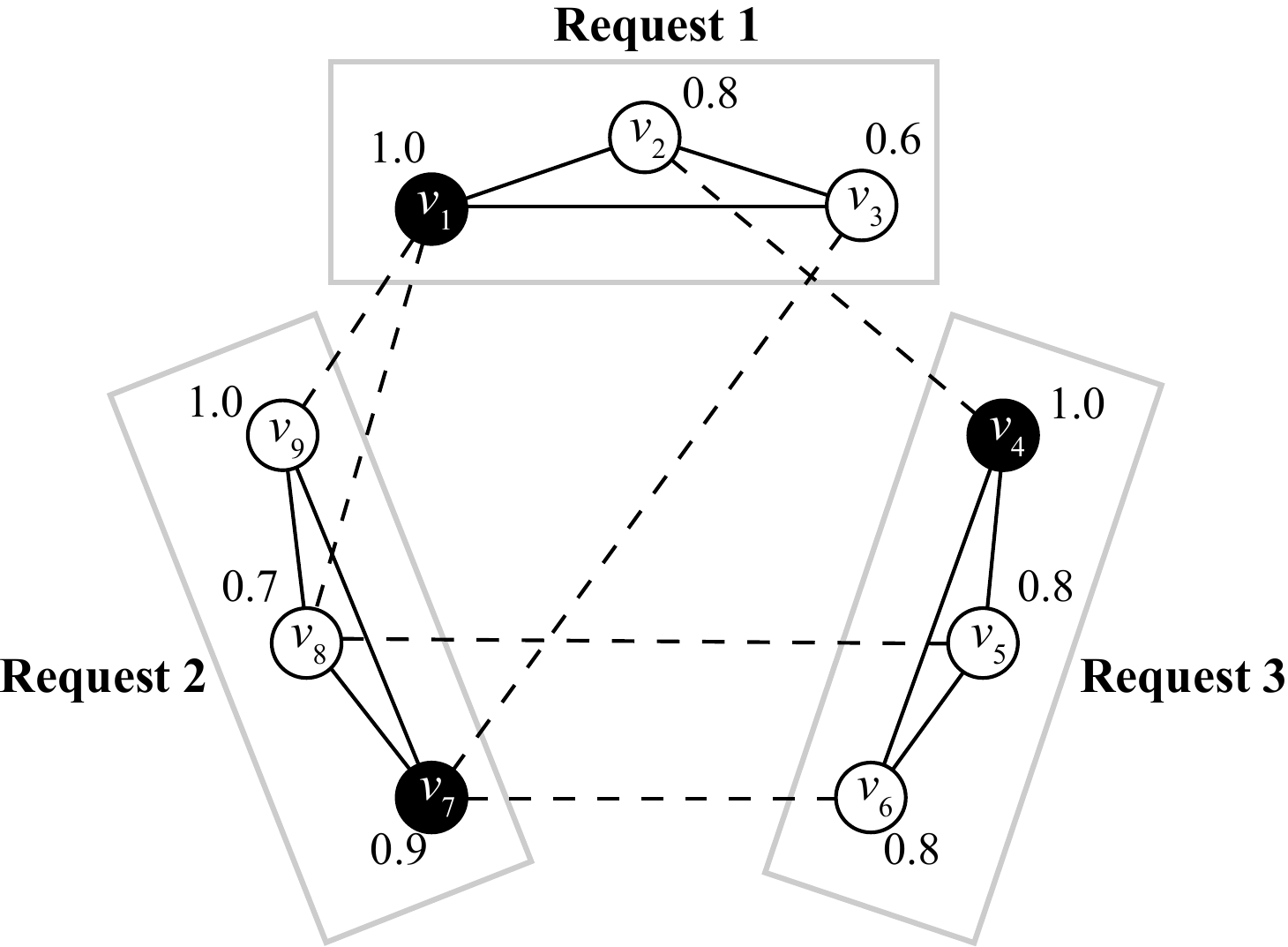}
    \caption{\textbf{The solution.} Vertices filled in black are selected as members of the independent set. The total weight of the selected vertices is 2.9, which is the maximum value in this case.}
    \label{fig:MWIS_graph_solution}
  \end{subfigure}
  \caption{An example of a generated graph and its corresponding solution to the MWIS problem.}
  \label{fig:MWIS_graph}
\end{figure}

We formulate the route selection process as an MWIS problem by constructing a graph to model the relationships between candidate routes.
We define the vertex set $V$ as the set of all candidate routes. 
For each request, edges are added between its candidate routes, forming a complete subgraph, ensuring that at most one vertex from each request can be included in the independent set.
Additionally, edges are introduced between any pair of candidate routes that are in conflict, guaranteeing that no conflicting routes are selected in the independent set.
Next, we define a weight $w_i$ for each vertex $i \in V$ as follows:
\begin{equation}
  w_i = \frac{d^*}{d_i}
\end{equation}
where $d_i$ is the length of the route corresponding to vertex $i$, and $d^*$ is the length of the shortest path connecting the source and destination of route $i$. By this definition, the weight equals 1 if the route is the shortest, and is smaller than 1 if the route is longer than the shortest path.
The objective is to maximize the total weight of the independent set, which simultaneously increases the number of approved requests and reduces the overall route lengths.
Note that maximizing the number of approved requests is not necessarily the primary objective, as we aim to avoid approving requests with excessively long routes.
For instance, selecting a single route with a weight of 1 is preferable to selecting two routes, each with a weight of less than 0.5.
The balance between maximizing the number of approved requests and minimizing route lengths can be adjusted by adding a constant offset to both the numerator and denominator of the weight definition $w_i$, though this refinement is not considered in this study.
An example of graph generation and its solution is illustrated in Fig. \ref{fig:MWIS_graph}.

By solving the MWIS problem on the constructed graph, we can identify the optimal combination of routes that meet the separation constraints.
However, due to the intractability of MWIS problems, solving them exactly or even approximating them efficiently is generally considered difficult \cite{trevisanInapproximabilityCombinatorialOptimization2014}. 
As a result, heuristic methods are commonly employed to address large-scale MWIS problems.
One such approach is the greedy algorithm, which, although it does not guarantee optimality, is computationally efficient. 
Several greedy algorithms have been explored for MWIS problems, and in this study, we adopt the method proposed by Sakai \textit{et al}. \cite{sakaiNoteGreedyAlgorithms2003b}. 
We describe the greedy algorithm for MWIS problems in Algorithm \ref{alg:greedy}.
This algorithm guarantees that the size of the independent set obtained is at least $\sum_{i \in V} w_i / (d(i) + 1)$.

\begin{algorithm}[tbp]
    \caption{Greedy algorithm for MWIS problems}
    \label{alg:greedy}
    \begin{algorithmic}
        \Require{A weighted graph $G$.}
        \Ensure{A maximal weighted independent set $I$.}
         \State $I \leftarrow \emptyset , i \leftarrow 0, G' \leftarrow G$;
         \While{$V (G') \neq \emptyset$}
           \State Select $i$ such that $\sum_{j \in N_{G'} (i)} w_j / \{d_{G'} (u)  +1\} \le w_i $
           \State $I \leftarrow I \cup \{i\}$;
           \State $G'\leftarrow G'[V(G') \setminus N_{G'} (i)]$
         \EndWhile
    \end{algorithmic}
\end{algorithm}

Next, we introduce quantum annealing as an optimization method for solving MWIS problems. 
To utilize quantum annealing for MWIS, the problem must first be reformulated as a quadratic unconstrained binary optimization (QUBO) problem.
A general QUBO problem is expressed as follows:
\begin{equation}
  \begin{split}
    \text{minimize} & ~~~~ \sum_{i,j} Q_{ij} x_i x_j \\
    \text{subject~to} & ~~~~ x_i \in \{0,1\} ~~~~ \forall i \in \{1, 2, \ldots ,N \}
  \end{split}
\end{equation}
where $Q$ is a coefficient matrix, $x_i$ represents a binary variable, and $N$ is the number of variables.
To handle constraints in QUBO problems, they are incorporated into the objective function as additional penalty terms.
These terms are designed to ensure that violations of the constraints incur a high cost, thus discouraging solutions that break them. 
By applying the penalty method to the original MWIS formulation \ref{eq:MWIS_IP}, we can remove the inequality constraints and reformulate the problem into the following QUBO form:
\begin{equation}
  \label{eq:MWIS_QUBO}
  \begin{split}
    \text{minimize} & ~~~~ - \sum_{i \in V}  w_i x_i + \lambda \sum_{ij \in E} x_i x_j \\
    \text{subject~to} & ~~~~ x_i \in \{0,1\} ~~~~ \forall i \in V.
  \end{split}
\end{equation}
The first term in equation \ref{eq:MWIS_QUBO} corresponds to the objective function of the MWIS problem, while the second term represents the penalty for violating the constraint. If a pair of adjacent vertices is selected in the independent set, i.e., if $x_i = x_j = 1$, the penalty $\lambda$ is added to the objective function. 
Here, $\lambda$ is the penalty coefficient, and by setting $\lambda$ sufficiently large, the solution to the Ising problem satisfies the constraints.
Given that the weights are defined to be less than or equal to 1, setting the penalty coefficient $\lambda$ to 2 is sufficient to ensure constraint satisfaction.

QUBO problems can be equivalently transformed into the problem of minimizing the energy of the Ising model, for which the Hamiltonian is expressed as:
\begin{equation}
  H_0(\mathbf{\sigma}) = - \sum_{i,j} J_{ij} \sigma_i \sigma_j - \sum_i h_i \sigma_i
\end{equation}
where $\sigma_i$ is a Ising spin variable that takes either $1$ or $-1$ and $J_{ij}$ and $h_i$ are the coupling constant between Ising spins and the bias term, respectively.
In quantum annealing, quantum fluctuations are leveraged to explore low-energy states of the QUBO problem.
The D-Wave quantum processor operates this process using superconducting qubits, with quantum fluctuations controlled by the strength of a transverse field.
The system’s Hamiltonian is expressed as:
\begin{equation}
  \hat{H}(s) = - A(s) \sum_i \hat{\sigma}_i^x + B(s) \hat{H}_0
\end{equation}
where $\hat{\sigma}_i^x$ represent the $x$-component of the Pauli matrices. 
The term $\hat{H}_0$ is the problem Hamiltonian, which encodes the Ising problem, and is obtained by replacing the Ising spin variables with the $z$-component of the Pauli matrices, $\hat{\sigma}_i^z$.
The system is governed by a predefined annealing schedule, parameterized by $0 \le s \le 1$.
The functions $A(s)$ and $B(s)$ are defined such that $A(0) \gg B(0)$ and $A(1) \ll B(1)$, ensuring that at $s = 0$, the system starts in a trivial ground state, where qubits are in a uniform superposition of all possible states.
As $s$ approaches 1, the system evolves into a nontrivial classical state, with spin variables that correspond to the solution of the QUBO problem.
The quantum adiabatic theorem ensures that if the schedule parameter $s$ is varied slowly enough, the system will remain in its ground state throughout the evolution.
Consequently, the optimal solution to the Ising problem should, in theory, be obtained at the conclusion of the annealing process.
In practice, however, the D-Wave quantum annealer does not perform ideal quantum annealing, and the optimization performance remains limited at the current stage of development. 
Despite this, the quantum annealer's ability to rapidly sample solutions can be effective for finding relatively good solutions, making it a useful heuristic solver. 
Moreover, the performance of quantum annealing hardware continues to improve.
This potential for growth is why we are particularly interested in quantum annealing as an optimization method from a long-term perspective.

\section*{Results}
\addcontentsline{toc}{section}{Results}
In this section, we present the results of routing and scheduling for a fleet of UAM vehicles using the OneSky's UTM simulator.

The simulator is customized for the airspace over Singapore, with the routing network constructed as depicted in Fig. \ref{fig:routing_network}.
Within this environment, we generate a set of requests and simulate the operations of a UAM fleet.
Requests are generated every 30 seconds by randomly selecting a pair of aerodromes in the routing network as the source and destination.
The desired start time for each request is set to zero, with an allowable delay of up to 60 seconds from the request generation time.
The total simulation time is 3500 seconds, with routing and scheduling updates performed at 30-second intervals.
The minimum required separation distance between two UAM vehicles is 100 meters.
We generate five candidate routes for each request to be evaluated during the scheduling and routing process.

The optimization computations are performed using an Intel Xeon Gold 6130 CPU with 141 GB of RAM. 
The greedy algorithm implemented in Python 3.8.8 is used for approximate solutions.
Gurobi Optimizer 9.1.2, a state-of-the-art mathematical optimization solver, is employed to obtain exact solutions.
Gurobi is widely recognized as the industry standard for solving complex optimization problems involving integer programming and is focused on delivering high-performance results.
Quantum annealing is carried out using the D-Wave Advantage 1.1 quantum processor.
The number of samples for the quantum annealing process is 100, with an annealing time of 1 $\mu$s.
To implement the QUBO problem \eqref{eq:MWIS_QUBO} on the D-Wave quantum processor, we employ a heuristic algorithm to map the problem graph onto the hardware graph using graph minor embedding techniques \cite{caiPracticalHeuristicFinding2014}.
After obtaining samples from the quantum processor, a post-processing step refines the solutions, ensuring they align with the original QUBO problem.
In cases where the values on redundant qubits representing the same logical variable differed, such as chain breaks, we applied a majority vote strategy to resolve the conflicts.
Chain breaks, caused by the limited connectivity between qubits in the current D-Wave annealer, often degrade the optimization's performance.
To mitigate this issue, we employed a steepest descent algorithm to refine the post-processed samples, further enhancing the quality of the solutions obtained from the quantum annealer.

\begin{figure}[tbp]
  \centering
  \includegraphics[width=13cm]{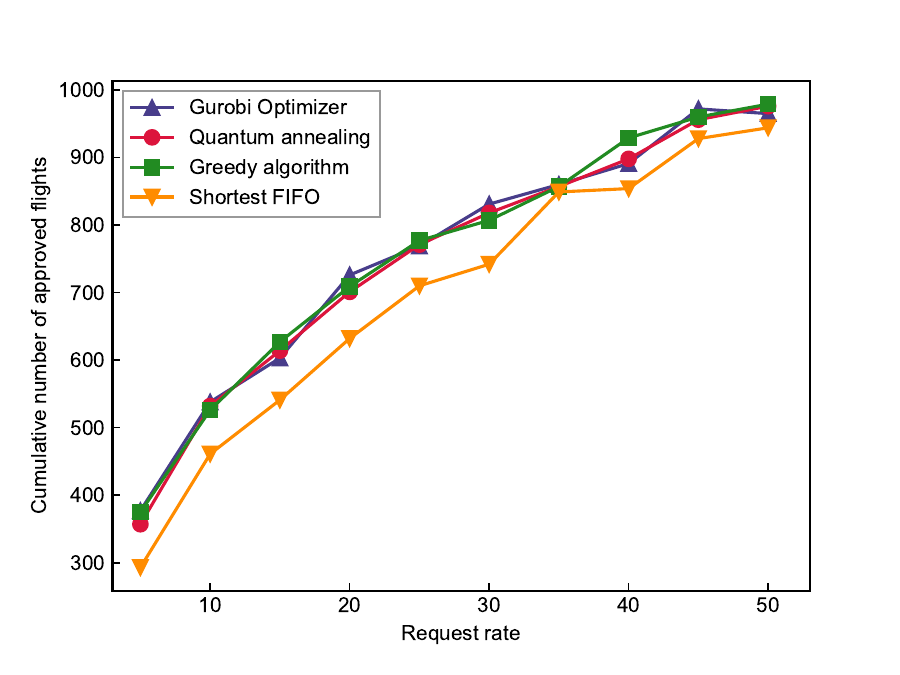}
  \caption{\textbf{Cumulative number of approved flights.} The plots show the cumulative number of approved flights during the simulation. The triangle, circle, square, and down-pointing triangle markers represent the results of the shortest FIFO, quantum annealing, exact optimization, and greedy algorithm, respectively.}
  \label{fig:approved_flights_number}
\end{figure}

\begin{figure}[tbp]
  \centering
  \includegraphics[width=13cm]{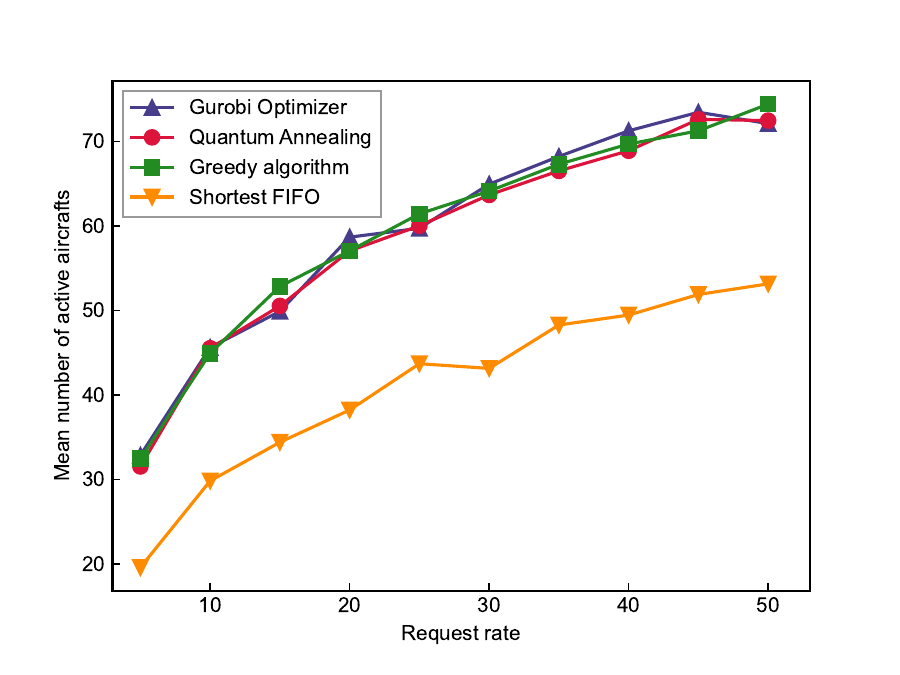}
  \caption{\textbf{Cumulative number of approved flights.} The plots show the average number of aircraft in the airspace at any given time during the simulation. The triangle, circle, square, and down-pointing triangle markers represent the results of the shortest FIFO, quantum annealing, exact optimization, and greedy algorithm, respectively.}
  \label{fig:active_aircrafts}
\end{figure}

We first evaluate our optimization framework by comparing several methods against a non-optimized baseline strategy, which we call the shortest first-in, first-out (FIFO) method.
The shortest FIFO method is a simple strategy that assigns each request the shortest available route and schedules flights in the order of arrival, provided the route is deemed safe.
Figure \ref{fig:approved_flights_number} illustrates the number of approved requests.
Our scheduling and routing framework approves more requests than the shortest FIFO method, demonstrating that route optimization in fleet management effectively increases the number of approved flights.
However, the difference between our approach and the shortest FIFO method diminishes when the request rate is high.
We believe that when the request rate is dense, the airspace becomes heavily congested, and the routing network reaches full capacity, leaving little room for further route adjustments.
We provide visualizations of the simulation results in Supplementary Movie 1.
In these visualizations, we observe that when using the shortest FIFO method, aircraft tend to utilize corridors with high centrality in the routing network, leading to congestion.
In contrast, our optimization approach mitigates this congestion by distributing the aircraft evenly throughout the airspace.
We also present the average number of active aircraft at any given time in Fig. \ref{fig:active_aircrafts}.
Our approach shows a more significant increase in active aircraft than the shortest FIFO method. 
This suggests that routes are more evenly distributed throughout the airspace, leading to more efficient utilization of available space. 
This implies that some routes may require detours to achieve the best combination of deconflicted paths, allowing for higher air traffic capacity while maintaining safety.

\begin{figure}[tbp]
  \centering
  \includegraphics[width=10cm]{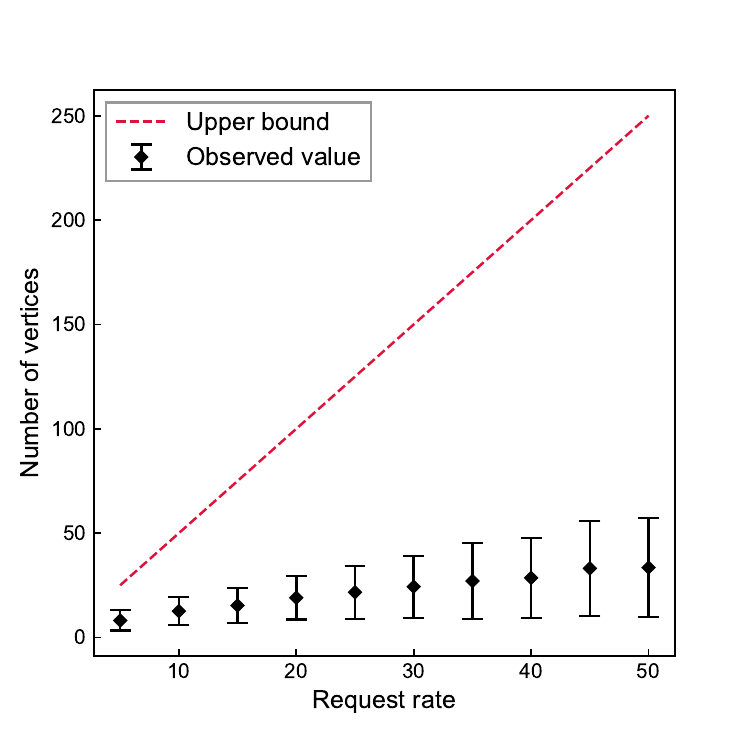}
  \caption{\textbf{Number of vertices.}  The plot shows the number of variables (vertices) in the MWIS problems at various request rates.}
  \label{fig:number_of_variables}
\end{figure}

The difference in performance between the various optimization schemes for MWIS problems is not significant in this case. 
To better understand the computational difficulty of the MWIS problems encountered during the simulation, we counted the number of variables involved in these problems, as illustrated in Fig. \ref{fig:number_of_variables}.
While the number of variables increases with the request rate, there is a large gap between the actual problem size and the theoretical upper bound.
This situation will likely occur when the number of incoming requests exceeds the capacity that can be efficiently managed through scheduling and routing. As the airspace is already at full capacity, there is little room for further optimization.
In such situations, the size of the MWIS problems does not grow substantially; thus, computational complexity is not a major concern.
Consequently, we conclude that each optimization algorithm can optimally solve most of the problems, which explains the minimal difference in performance observed under the current routing network capacity.
However, if the routing network's capacity were expanded, such as by increasing the vertical dimension or using other methods, the number of variables in the MWIS problems would increase significantly.
As a result, the computational hardness of the MWIS problems would become a more critical factor, and the performance of the optimization algorithms would play a much larger role in determining overall effectiveness.

Finally, we evaluate the current performance of each optimization method in terms of both computation time and solution quality.
Since quantum annealing is a stochastic algorithm, it is appropriate to run the algorithm multiple times and measure the probability of obtaining the optimal solution within a given time rather than solely focusing on the average runtime.
To assess this, we employ the time-to-solution (TTS) benchmark as a performance metric, which is defined as:
\begin{equation}
  \text{TTS}(p) = t_c \frac{\log (1-p)}{\log (1-p_{\text{opt}})}
\end{equation}
where $p$ represents the probability of obtaining the optimal solution at least once after a fixed number of trials,  $p_{\text{opt}}$ is the probability of obtaining the optimal solution in a single trial, and $t_c$ is the computational time for a single trial. 
For instance, $\text{TTS}(0.99)$ indicates the estimated time required to achieve the optimal solution with a 99\% probability.
We extracted 629 MWIS problems from the UTM simulation in Singapore, with problem sizes ranging from 5 to 50 variables, as illustrated in Fig. \ref{fig:number_of_variables}, and evaluated the performance of each algorithm.
We set the number of samples for quantum annealing to 10,000, with an annealing time of 1 $\mu$s per trial.

Understanding the operation and timing of D-Wave's quantum processor is crucial to measuring the runtime of quantum annealing.
Defining the runtime for a single trial as the annealing time makes sense for evaluating the pure performance of quantum annealing.
However, in practice, the current D-Wave quantum processor requires additional computation processes beyond annealing, meaning that the annealing time alone does not fully determine the runtime.
Briefly, the runtime of the quantum annealing is divided into programming time and sampling time, collectively referred to as quantum processor unit (QPU) access time.
Programming time refers to configuring the quantum processor with the problem, representing a one-time overhead independent of the number of samples.
On the other hand, sampling time refers to obtaining samples from the quantum processor and is proportional to the number of samples.
It includes the annealing time, post-processing time for readout, and the delay required to reinitialize the processor.
While programming time and post-processing are nonessential for the theoretical performance of quantum annealing, they are critical in practical use.
Therefore, we evaluate the total annealing time and QPU access time.
This analysis does not consider other overheads, such as communication time and pre- or post-processing on the classical computer.

\begin{figure}[tbp]
  \centering
  \includegraphics[width=13cm]{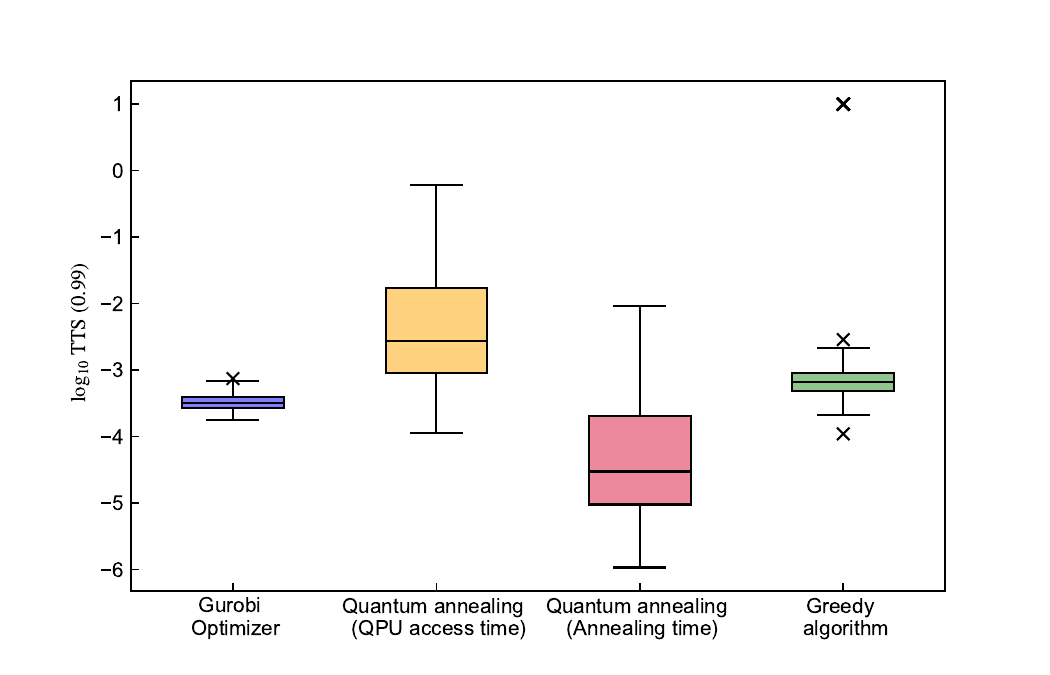}
  \caption{\textbf{TTS benchmark.} The box plots show the time-to-solution benchmark of quantum annealing, Gurobi Optimizer, and the greedy algorithm. The box plots represent the distribution of the TTS values and are drawn from the first to the third quartiles. The whiskers are drawn 1.5 times the interquartile range from the first and third quartiles. Outliers are shown with cross markers.}
  \label{fig:tts}
\end{figure}

For the Gurobi Optimizer and the greedy algorithm, we use the runtime of a single trial rather than time-to-solution (TTS).
As the greedy algorithm is deterministic and does not guarantee the optimal solution, we set the runtime to a fixed 10 seconds when the optimal solution is not obtained.
The results are illustrated in Fig. \ref{fig:tts}.
Quantum annealing outperforms Gurobi Optimizer in terms of annealing time.
However, the opposite is true when considering the total QPU access time.
Both the Gurobi Optimizer and the greedy algorithm show superior consistency in runtime, with quantum annealing exhibiting greater variance across different instances.
The greedy algorithm appears slower than Gurobi Optimizer in some cases.
We attribute this to our current implementation of the greedy algorithm, which is likely not fully optimized, leaving room for performance improvements.
Additionally, a few problems were not solved optimally by the greedy algorithm.
Nevertheless, all the algorithms managed to solve most of the problems optimally within 0.1 seconds, with the differences in performance being relatively minor for the current simulation conditions.
That said, the difference between solvers is expected to become more pronounced as the routing network capacity increases and the problem size grows.
In this context, it is noteworthy that the current D-Wave quantum annealing machine performs competitively with classical solvers, even when accounting for QPU access time.
Furthermore, quantum annealing could become even more competitive as its performance potentially improves over the long term.

\section*{Discussion}
\addcontentsline{toc}{section}{Discussion}
In this study, we proposed a path planning and scheduling framework for the fleet of UAMs in urban airspace.
The routing decision was formulated as a MWIS problem, allowing us to leverage existing optimization algorithms, including quantum annealing. 
We evaluated the performance of the framework using the OneSky UTM simulator for the Singapore airspace and compared the results with the shortest FIFO method.
The results indicate that the proposed framework is effective when the request rate is moderate.
We further compared the performance of each optimization algorithm in terms of runtime and solution quality using the TTS benchmark.
The findings show that all the algorithms could provide fast solutions, making them reasonable options for our scheduling and routing framework.
The comparison between quantum annealing and classical solvers reveals that the current D-Wave machine is competitive with traditional methods.
Moreover, when considering only annealing time, quantum annealing outperforms classical solvers, although the problem sizes in our study were relatively small.
This suggests that quantum annealing has the potential to be a powerful tool for solving hard combinatorial optimization problems in the future, especially as its performance continues to improve.

As future work, extending our framework to optimize the timing of flight departures and arrivals presents a promising direction.
In this study, flights were scheduled greedily to approve as many requests as possible at each time step without optimizing the departure and arrival times.
This strategy can lead to suboptimal outcomes over time, potentially resulting in airspace congestion.
To mitigate this, optimizing the timing of departures and arrivals, alongside flight approval and routing decisions, will be crucial to prevent such congestion and improve overall efficiency.

We are also interested in simulating larger-scale urban airspace.
The Singapore airspace used in this study is relatively small, and the capacity of the routing network was limited when assessing the problem size.
In larger urban airspaces, the capacity of the routing network is likely to increase, making the impact of optimization more significant.
For example, expanding the routing network into three-dimensional space is one approach to increase its capacity and improve traffic efficiency.
In such scenarios, the computational performance of optimization methods becomes crucial, and next-generation algorithms, such as quantum annealing, may offer a viable solution for tackling these larger, more complex problems.
At the same time, managing airspace for UAM vehicles presents a challenging issue for aviation authorities and service providers. We believe that advancing aerial transportation through UAM vehicles will drive the need for global optimization in airspace management, and our framework could contribute to the overall solution.

This study identified quantum annealing as a promising candidate for solving combinatorial optimization problems.
To utilize quantum annealing, problems must first be encoded as QUBO formulations, and the effectiveness of this formulation directly influences the performance of the quantum annealing process.
We successfully reduced the routing and scheduling problems to the well-known MWIS problem, enabling the application of already established optimization algorithms.
Additionally, in the context of quantum annealing, further performance improvements will depend on developing specialized methods tailored to specific problem types, as its current capabilities for solving general QUBO problems remain limited.
The generality of typical combinatorial problems, like MWIS, broadens the applicability of quantum annealing to real-world challenges.
Thus, exploring specialized optimization techniques using quantum annealing for typical combinatorial problems is an intriguing direction. For instance, graph coloring problems and capacitated vehicle routing problems are examples where specialized approaches have proven effective \cite{feldHybridSolutionMethod2019a, kwokGraphColoringQuantum2020}.
However, the development of specialized methods for quantum annealing is still in its early stages, making future research in this area highly promising.

\bibliography{citation}

\section*{Acknowledgements}
\addcontentsline{toc}{section}{Acknowledgements}
The authors would like to express their gratitude to Daniel Honaker and Tim Carrico from OneSky Systems for their invaluable contributions in providing the UTM simulation environment and animated visualizations of UAM systems.
This study was financially supported by programs for bridging the gap between R\&D and IDeal society (Society 5.0) and Generating Economic and social value (BRIDGE) and Cross-ministerial Strategic Innovation Promotion Program (SIP) from the Cabinet Office.

\section*{Author contributions statement}
\addcontentsline{toc}{section}{Author contributions statement}
R.H. conceived of the presented idea and performed the experiments.
T.M., R.U., G.E.,and K.T. helped to supervise the project.
M.T. contributed to the management of the research project.
M.O. verified the analytical methods and supervised the findings of this work. 
All authors discussed the results and contributed to the final manuscript.

\section*{Additional information}
\addcontentsline{toc}{section}{Additional information}
\textbf{Competing interests}: The authors declare no competing interests. \vspace{1ex} \\ 
\textbf{Correspondence} should be addressed to R.H.  \vspace{1ex} \\ 
\textbf{Data availability}: The datasets used during the current study are available from the corresponding author upon reasonable request.

\end{document}